\documentclass[pra,aps,showpacs,twocolumn,groupedaddress,superscriptaddress,
floatfix]{revtex4}
\usepackage{epsfig}
\usepackage{graphicx}
\usepackage{amsfonts}
\usepackage{amsmath}

\begin{document}
\title{Two-photon interference with true thermal light}
\author{Yan-Hua Zhai$^1$, Xi-Hao Chen$^{1,2}$, Da Zhang$^1$, Ling-An Wu}
\thanks{Corresponding author: wula@aphy.iphy.ac.cn}
\affiliation{Institute of Physics, CAS, Beijing 100080,
China\\\it{$^2$}Department of Physics, Liaoning University,
Shenyang 110036, China}

\begin{abstract} Two-photon interference and "ghost" imaging with
entangled light have attracted much attention since the last
century because of the novel features such as non-locality and
sub-wavelength effect. Recently, it has been found  that
pseudo-thermal light can mimic certain effects of entangled light.
We report here the first observation of two-photon interference
with true thermal light.
\end{abstract}


 \maketitle
Interference is described in textbooks as the coherent
superposition of waves, and the ability to interfere is regarded
as a special attribute, known as coherence, of the radiation
source. Optical coherent sources include lasers high above
threshold, but the majority of light sources such as thermal light
are incoherent. To describe the intensity of the field from
coherent sources the amplitudes from all sources present are
superposed and summed together before the intensity is calculated,
whereas in the case of incoherent sources, the intensities from
all the sources can generally be added directly. Apart from the
first order field intensities, different sources exhibit different
characteristics in their higher order intensity properties, and
this has led to widespread studies of the nature of a variety of
quantum and classical, coherent and incoherent light sources. Many
interesting and important applications have been developed
therefrom.

Of particular interest are the second order intensity
characteristics, which of course are easier to investigate than
the higher orders. Since Hanbury-Brown and Twiss (HBT) \cite{hbt}
first measured the joint-intensity of light, coincidence
measurements have been applied to measure the second-order
coherence function($G^{(2)}$) of various photon fields. In
particular, novel phenomena such as ``ghost" imaging \cite{shih1}
and ``ghost" interference \cite{shih2} have been observed through
coincidence measurements of the light generated by spontaneous
parametric down-conversion (SPDC). In the ghost interference
experiment with an SPDC entangled light source, the two-photon
amplitudes from both slits are added together as with coherent
light sources, although the field would seem to be incoherent from
the usual viewpoint. Related to these kinds of phenomena are
experiments that demonstrate seemingly nonclassical effects such
as sub-wavelength diffraction \cite{shih3}. At one time it was
thought that these phenomena were exclusive to quantum entangled
light \cite{ayman}. However, it has recently been shown
theoretically that thermal light can generate similar effects to
those of entangled light \cite{kaige,lugiato1,zhu,han}, and
experiments to prove this have been performed using pseudo-thermal
light \cite{shih4,jun, lugiato2}. However, in all these
experiments the primary light source was a He-Ne laser, the
coherent beam being converted to pseudo-thermal light by a
rotating ground glass plate or some other means. Different from
these previous investigations we report the first two-photon
sub-wavelength experiment with true thermal light (sometimes
called chaotic light), an incoherent light source exhibiting
thermal statistics that cannot interfere in the conventional sense
of the term, i.e. does not exhibit first order interference.

We should mention here that the distinction between thermal and
chaotic light is not so well defined and different authors give
different definitions. In Ref. \cite{Goodman}, thermal light is
described as the radiation emitted by spontaneous emission from
``a large collection of atoms or molecules, excited to high energy
states by thermal, electrical, or other means" when they``randomly
and independently drop to lower energy states", and the sun,
incandescent bulbs and gas discharge lamps are cited as examples.
Some authors specify that discharge and filament lamps as well as
thermal cavities are forms of chaotic light sources \cite{Loudon,
Bachor}, while thermal light is the broad spectral emission from a
thermally excited glowing filament \cite{Bachor}. Mandel and Wolf
define thermal radiation as ``radiation that is derivable from
blackbody radiation by any linear filtering process" and point out
that ``It has sometimes also been called chaotic radiation"
\cite{Mandel}. Since a hollow cathode lamp can be portrayed by
thermal statistics, we employ the term ``thermal light" here in
order to correspond and contrast with previous two-photon
interference experiments that used pseudo-thermal light.

\begin{figure}
\includegraphics[width=7cm]{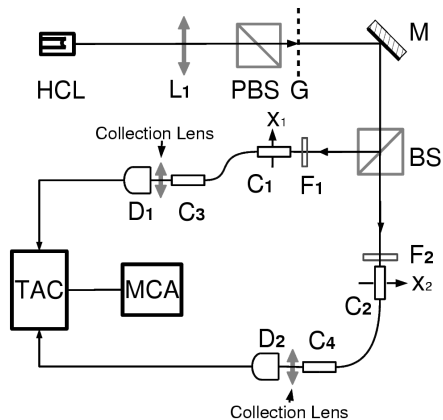}
\caption{Experimental set-up. HCL: hollow cathode lamp;
$\mathrm{L_1}$: lens of focal length 10cm; G: grating of groove
width 0.08mm and groove spacing 0.2mm; $\mathrm{F_1,F_2}$:
interference filters transmitting about $70\%$ at 780nm. PBS:
polarizing beam splitter; BS: non-polarizing beam splitter;
$\mathrm{C_1}$ - $\mathrm{C_4}$: fiber collimators. Effective
diameter of fiber collimators in front of the detectors
 is 2mm. }
\label{setup}
\end{figure}

An outline of the experimental set-up is shown in Fig.
\ref{setup}. We employed a commercial rubidium hollow-cathode lamp
\cite{Bernhard} manufactured by the General Research Institute for
Nonferrous Metals (China), which is the type commonly used in
atomic absorption spectroscopy because of its relatively sharp
spectral line width. The lamp was powered by a direct current of
20mA in our experiments, and the resonance wavelength was 780nm.
The coherence time $\tau_0$ was estimated from an HBT type
measurement of the second-order correlation function to be about
0.2ns~\cite{image}, which is much shorter than that of previous
experiments using randomly scattered light from a He-Ne laser.

In Fig.~\ref{setup} the light from the lamp is focused by the
convex lens ($\mathrm{L_1}$) of 10cm focal length  onto a
diffraction grating (G) to form a secondary light source. A
polarizing beam splitter (PBS) just before the grating allows only
the  horizontally polarized component of the incoherent beam to
pass. The inner diameter of the hollow cathode is 3mm, from which
its image on the grating is calculated to be about 1mm in
diameter. The width $b$ of a grating groove is 0.08mm and the
spacing  between grooves $d=0.2$mm, so five slits are illuminated.
After reflection by a mirror (M) the beam is divided by a
$50\%/50\%$ non-polarizing beam splitter (BS). The reflected and
transmitted beams pass through interference filters $\mathrm{F_1}$
and $\mathrm{F_2}$ before being coupled into single photon
detectors $\mathrm{D_1}$ and $\mathrm{D_2}$ (Perkin Elmer
SPCM-AQR-13), respectively, through fiber collimators
$\mathrm{C_1}$ and $\mathrm{C_2}$, and finally collection lenses.
Both collimators $\mathrm{C_1}$ and $\mathrm{C_2}$ can be
translated horizontally across the beam. The transmission of the
interference filters is about $70\%$ at 780nm and the receiving
lens of the collimators about 2mm in diameter. The distance $z$
from the grating(G) to either detector is 162cm. The detector
output signals are sent to a time-amplitude converter (TAC), with
$\mathrm{D_1}$ and $\mathrm{D_2}$ providing the ``start'' and
``stop'' signals, respectively. The TAC output is connected to a
multi-channel analyzer (MCA), and the computer displays a
histogram of the different intervals between the times of arrival
of the photons at the two detectors. From this we obtain the
relation between the photon count rate and time interval $\tau$,
and subsequently the second-order correlation function

\begin{eqnarray}  \label{G2}
 G^{(2)}(\tau) = \langle \hat{E}_{2}(\tau)^{(-)}
 \hat{E}_{1}(0)^{(-)}
 \hat{E}_{1}(0)^{(+)} \hat{E}_{2}(\tau)^{(+)}  \rangle,
\end{eqnarray}
where $\hat{E}_{i}^{(+)}$, $\hat{E}_{i}^{(-)}$ are the positive
and negative frequency field operators at detectors $\mathrm{D}_i
(i=1,2)$, respectively.

\begin{figure}
 \includegraphics[width=7cm]{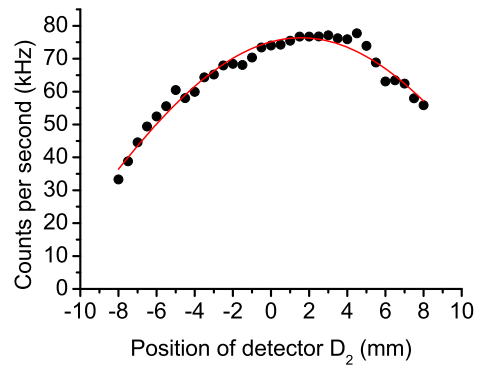}
\caption{Single detector counts vs. position of detector
$\mathrm{D_2}$. The solid curve is a Gaussian fit of data points.}
 \label{singlecount}
\end{figure}

To begin with, the detector $\mathrm{D_1}$ was kept fixed while
$\mathrm{D_2}$ (collimator $\mathrm{C_2}$) was scanned in the
horizontal direction and the single counts of $\mathrm{D_2}$
recorded as a function of its position. As can be seen from Fig.
\ref{singlecount}, no first-order interference pattern was
observed so there is no first-order coherence in our experiment.

\begin{figure}
 \includegraphics[width=7cm]{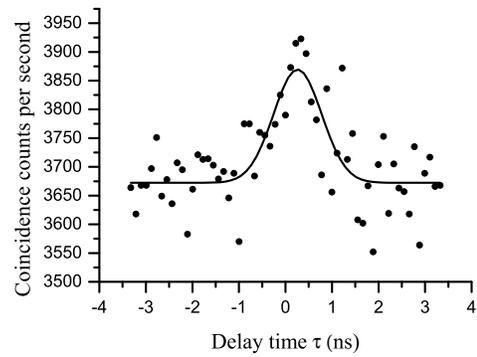}
\caption{Joint detection counts vs time interval of the photons
detected at $\mathrm{D_1}$ and $\mathrm{D_2}$. The solid curve is
a Gaussian fit of data points.}
 \label{si1300}
\end{figure}

Next, the collimators $\mathrm{C_1}$ and $\mathrm{C_2}$ were both
fixed in the center of their beams at the positions $x_1=0$,
$x_2=0$, respectively. The distribution of the times of arrival of
photons at the detectors as displayed on the MCA is shown in Fig.
\ref{si1300}, from which we obtained the value of $g^{(2)} $
\cite{ image} by dividing the values of the average
$G^{(2)}(t_2-t_1)$ for $|\tau| \leq 0.25ns$ (corresponding to
signals arriving almost simultaneously) by the value of $G^{(2)}(
\tau)$ for $|\tau| \gg 1.3ns$ (corresponding to signals arriving
randomly well beyond any correlation times), i.e.

\begin{eqnarray}
g^{(2)}
=\frac{G^{(2)}(|\tau|\leq0.25\mathrm{ns})}{G^{(2)}(|\tau|\geq1.3\mathrm{ns})}
.
\end{eqnarray}

\begin{figure}
\includegraphics[width=7cm]{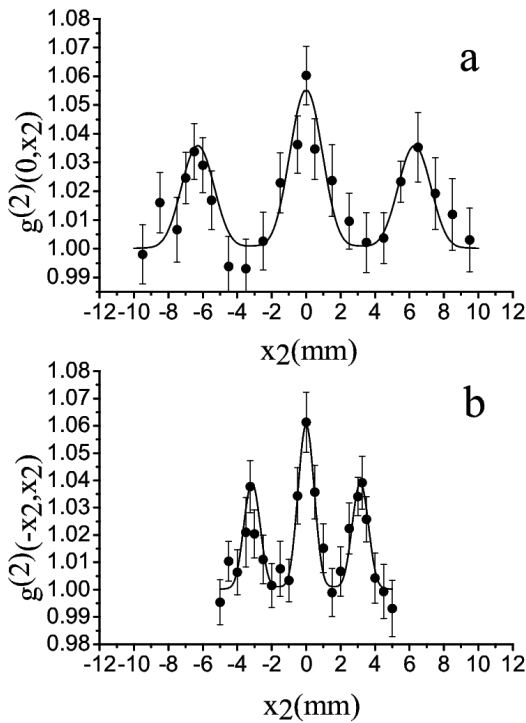}
\caption{ Normalized second-order correlation function of thermal
light (a) $g^{(2)}(0,x_2)$ vs position of detector $\mathrm{D_2}$
with detector $\mathrm{D_1}$ fixed. (b) $g^{(2)}(-x_2, x_2)$ vs
position of detector $\mathrm{ D_2}$. The solid curve is
calculated taking into consideration the finite size of the
detectors~\cite{hbt2}}.\label{x-x}
\end{figure}

With $\mathrm{C_1}$ still fixed at $x_1=0$, $\mathrm{C_2}$ was
then moved in steps of 0.5mm or 1mm through $x_2=\pm 10\mathrm{
mm}$. From the data obtained on the MCA the normalized
second-order correlation function $g^{(2)}$ was calculated, and
the second-order interference-diffraction pattern of the grating
plotted, as shown in Fig. \ref{x-x}a. We see that the pattern
looks classical, with a distance of about 6.5mm between the zero
and first order interference peaks. This agrees well with the
calculated value of 6.3mm obtained from the grating equation for
first-order interference-diffraction of the field seen at a single
detector \cite{wolf}.

\begin{eqnarray}
\sin{\theta}-\sin{\theta_0} = m\lambda/d \hspace{0.3cm}(m=0, \pm1,
\pm2, \cdots),
\end{eqnarray}
where $\theta_0$ is the angle of the incident light and $\theta$
that of the diffracted light measured from the normal to the
grating plane. The integer $m$ represents the path difference in
wavelengths between light diffracted in the direction of the
maximum, from corresponding points in two neighboring grooves.

\begin{figure}
 \includegraphics[width=7cm]{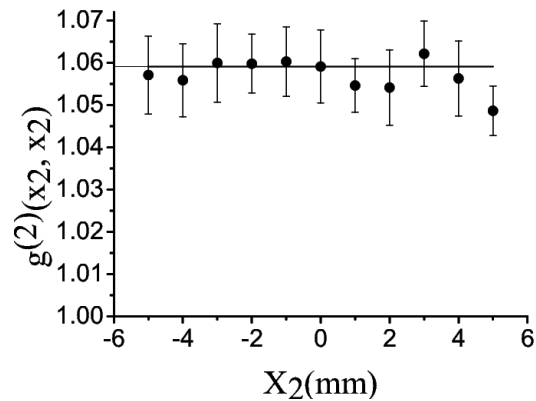}
\caption{Normalized second-order correlation function
$g^{(2)}(x_2, x_2)$ of thermal light vs position of detector
$\mathrm{D_2}$.}
 \label{xx}
\end{figure}

When the fiber collimators $\mathrm{C_1}$ and $\mathrm{C_2}$ were
scanned in \emph{opposite} directions $(x, -x)$ in steps of 0.25mm
or 0.5mm simultaneously, the second-order interference-diffraction
pattern of the grating shown in Fig. \ref{x-x}b was obtained. The
distance from the zero order to the first order interference peak
is about 3.25mm, which is exactly  half that of the classical
case. This is the well known sub-wavelength effect first predicted
and observed for two-photon interference with entangled photon
pairs \cite{shih3}.

However, when both collimators were scanned simultaneously in the
 \emph{same} direction $(x, x)$, no interference pattern
was observed, as shown in Fig. \ref{xx}. This is different from
the case with an entangled light source \cite{shih3}.

In our experiment, because the coherence time of the thermal light
source is shorter than the time resolution of the detection system
which is about 1ns, and the detectors are not point-like, the
maximum of $g^{(2)}$ cannot reach 2 and the visibility is only
about $3\%$ \cite{image,rubin}.

These experimental results are in good accordance with the values
predicted theoretically \cite{kaige,lugiato1,zhu} for the
detectors scanned in opposite directions,

\begin{eqnarray}
g^{(2)}(x,-x)-1\propto \mathrm{sinc^2}[\frac{\pi
bx}{(\lambda/2)z}] \mathrm{cos^2}[\frac{\pi dx}{(\lambda/2)z}]
\end{eqnarray}

and for the detectors scanned in the same direction,
\begin{eqnarray}
g^{(2)}(x,x)\propto \mathrm{const}.
\end{eqnarray}

To summarize, two-photon interference with sub-wavelength fringes
has been observed for the first time with true thermal light.
Although the visibility is low compared with entangled two-photon
interference, which exhibits high visibility but low intensity,
thermal light is of course much easier to generate and measure
than entangled light. The constant background in the
interference-diffraction pattern, an unavoidable feature of
thermal light sources, could nevertheless be removed by some
means, e.g. digitally.

It is interesting to note that for incoherent light sources that
do not exhibit first order interference in the plane of detection,
we can generally just sum the intensities produced by the
individual sources, instead of having to sum the field amplitudes.
However, for second order correlation measurements of the field at
two distinct space-time locations, we cannot merely add the
intensities even for incoherent thermal fields. It is the
individual field amplitude components that have to be summed, and
it is this that gives rise to the intensity product term of the
HBT experiment, and two-photon interference features. What is the
difference, or similarity, between classical  and entangled light
sources? It is evident that the two-photon interference and
imaging effects observed so far both originate in the correlation
of the photons arriving at the detectors. For entangled sources
the strong one-to-one correspondence of the two photons within
each pair allows a theoretical visibility of $100\%$ for both
ghost imaging and ghost interference. For classical thermal
sources it is the bunching effect that produces the weak but
finite correlation. These are all consequences of the photon
statistics, which can give rise to many rich and varied phenomena.

We are grateful to Yan-Hua Shih, De-Zhong Cao, Kai-Ge Wang, and
Shi-Yao Zhu for useful discussions. We also thank Zhan-Chun Zuo
and Hai-Qiang Ma for their experimental assistance. This work was
supported by the National Natural Science Foundation of China
(Grant No. 60178013), the National Program for Basic Research in
China (Grant 001CB309301), and the Knowledge Innovation Program of
the Chinese Academy of Sciences.

\end{document}